# Systematic grid-wise radon concentration measurements and first radon map in Cyprus

**G. Theodoulou[a], Y. Parpottas[b], and H. Tsertos[a]**
[a] Department of Physics, University of Cyprus, 1678 Nicosia, Cyprus
[b] General Department (Physics-Mathematics), School of Engineering and Applied Sciences, Frederick University, 1036 Nicosia, Cyprus

## Abstract

A systematic study of the indoor airborne radon concentration in the central part of the Nicosia district was conducted, using high-sensitivity active radon portable detectors of the type “RADIM3A”. From a total of 108 measurements in 54 grids of 1 $km^2$ area each, the overall mean value is 20.6±13.2 $Bq\ m^{-3}$ (A.M.±S.D.). That is almost twice less than the corresponding average worldwide value. The radon concentration levels in drinking water were also measured in 24 sites of the residential district, using the high-sensitivity radon detector of the type “RADIM3W”. The mean value obtained from these measurements is 243.8±224.8 $mBq\ L^{-1}$, which is relatively low compared to the corresponding internationally accepted level. The associated annual effective dose rates to each measurement were also calculated and compared to the corresponding worldwide values. From the geographical coordinates of the measuring sites and the corresponding radon concentration values, the digital radon map of the central part of the Nicosia district was constructed for the first time, by means of the ArcMap software package.

## 1. Introduction

Radon ($^{222}Rn$) is a radioactive inert gas produced in the naturally occurring Uranium ($^{238}U$) decay series, following the alpha decay of the Radium isotope ($^{226}Ra$). It directly decays into Polonium–218 ($^{218}Po$) by emitting alpha particles with energy of about 5.5 *MeV*. The amount of $^{238}U$ and $^{226}Ra$ isotopes, present in the underlying soils and in the building materials, determines the radon release in a house. Thus, the magnitude of $^{222}Rn$ concentration indoors depends primarily on the construction of the building and the amount of $^{222}Rn$ in the underlying soil. Daughter nuclides following radon decay, attached to microscopic dust’s particles, are inhaled and emit alpha particles, which effectively cause biological damages to the lung cells.

Therefore, inhalation of air with high radon concentration over a long period of time increases the risk of lung cancer.

It is quite clear that comprehensive measurements to determine the $^{238}$U and $^{226}$Ra isotopes in soils and building materials (see e.g. Tzortzis et al., 2004, Michael et.al., 2011) are very important for a complete understanding of radon activity. Towards this end, direct studies of radon release from the soil, similar to those conducted in other areas in the world (see e.g. Giammanco et al., 2009, Neri et al., 2011), are complementary and also very useful.

Drinking water can also be a significant source of radon. It can escape from household water to the indoor air by concentration ratio of $10^{-4}$ (UNSCEAR, 2000) or it can deliver a whole body radiation dose by ingestion of drinking water with a risk of stomach and colon cancer. The World Health Organization (WHO, 2005) and the U.S. Department of Health and Human Services (HHS, 2010) categorized the radon among the high-risk substances for carcinogenesis, after smoking.

The International Commission on Radiological Protection (ICRP, 2009), the International Atomic Energy Agency (IAEA, 2003), the European Commission (EC, 1997), and the United Nations Scientific Committee on the Effects of Atomic Radiation (UNSCEAR, 2000) refer to the health hazards due to radon inhalation, and emphasize the need each country to define upper limits for the radon concentration in old and new building, as well as to construct the associated radon maps. The defined upper limits for indoor radon concentration in most European countries are 200 $Bq\ m^{-3}$ for new buildings, and 400 $Bq\ m^{-3}$ for old buildings (EC, 1997). Many countries already constructed or are under construction of their own radon maps. Various mapping techniques and site selection criteria, such as districts or geological background or grid areas, were used for the construction of these maps (Synnott and Fenton, 2005). A lot of these maps are published online to inform individual or interested parties.

In 1993, Christofides and Christodoulides carried out the first measurements on indoor $^{222}$Rn concentration in Cyprus houses. Passive track detectors based on the CR-39 plastic element were utilized by these investigations (Christofides and Christodoulides, 1993). In 2003, the Nuclear Physics Laboratory of the Department of Physics, University of Cyprus, has begun the "Cyprus Radioisotopes Project", aiming to systematically register the indoor $^{222}$Rn concentration in Cyprus buildings and dwellings (Anastasiou et al., 2003). Other part of the project included extensive measurements of the naturally occurring radioisotopes in rock, soil, and building materials (Tzortzis et al., 2003a; Tzortzis et al., 2003b; Tzortzis et al., 2004; Tzortzis and Tsertos, 2004; Svoukis and Tsertos, 2007). These first survey measurements on indoor radon concentration were carried out using high-sensitivity active portable radon monitors of the type "RADIM3A" (Plch, 2001) described below. They directly determined the radon concentration in air in pre-selected sampling time intervals, at good statistical accuracy, within only a few hours of measurements. This is of great advantage, since the results are obtained immediately without sending the collection material to a laboratory for reading and further evaluation as it is the case by the passive devices.

An overall arithmetic mean value of 19.3±14.7 $Bq\ m^{-3}$ was determined from these survey measurements, throughout the whole part of the Republic of Cyprus (Anastasiou et al., 2003). As a next step, our laboratory has started systematic grid-wise measurements on airborne radon concentration throughout the main residential districts of Cyprus, using a constant grid area of 1 $km^2$ and high-sensitivity active radon portable detectors. In addition, radon concentration levels in drinking water were also measured. In this paper, the detection systems and methodology are described in detail. The first indoor airborne radon activity concentration results, together with the corresponding first digital radon map, for the high-population central part of the Nicosia district, are presented and discussed. Moreover, the radon levels of the drinking water in the district are also measured and reported. Last, the associated annual effective dose rates to the population are calculated and compared to the world average values.

## 2. Methodology

### *2.1. Detection*

The indoor measurements were carried out using two high-sensitivity active portable radon monitors. The RADIM3A (commercial name) is a compact and dedicated detector system designed to directly monitor the radon concentrations, determine the radon entry rate and ventilation coefficient. It incorporates additional sensors to simultaneously measure the pressure, temperature and relative humidity. The sampling time can be adjusted from 0.5 – 24 hours and therefore we can observe graphically any possible fluctuations regarding the radon concentration, pressure, temperature and relative humidity over the detection time (one record per sampling time). Also, the system corrects the results for the effect of humidity.

Moreover, calibration over the whole dynamic range of the instrument is made by the manufacturer (in relation to a reference RADM3A instrument) and the accuracy of the calibration is then verified in the State Metrological Institute of the Czech Republic. Verification is achieved by comparing the results of the instrument and the reference instrument using a secondary ATMOS standard (the ATMOS User Centre is a service of the German Agency, DFD) and then the calibration factor value is adjusted so as the tested instrument yields the same results as ATMOS. The overall uncertainty in the calibration procedure is then equal to 10% since the response of the ATMOS reference instrument is known with an error of about 7% (Plch, 2001).

The instrument is shown schematically in Fig. 1. The radon diffuses into the instrument where a filter collects all airborne radon decay products. The radon concentration is determined by measuring the α–activity of $^{218}Po$ (the energy of the α–peak appears at ~ 6 *MeV*), which is collected by the electric field on the surface of a semiconductor (Si) detector in the chamber of the detection system (Hopke, 1989). The chamber has an optimized spherical geometry half of which is formed by a grid covered with two layers of a dense cloth. This cloth prevents the entrance of airborne radon decay products and protects the detector against dust and light. Because most of the radon decay products are positively charged, the vessel is connected to the positive pole of a high–voltage supply and the surface of the Si detector electrode is connected to the negative pole. The positively charged $^{222}Rn$ decay products are neutralized by water vapor and other admixtures, and this effect is kept low by employing the highest possible electric field. The instrument uses a stabilized voltage power supply of 2 *kV,* which is controlled by the internal computer (see Fig. 1). The optimal form of the chamber and the high electric field reduces the influence of humidity.

The instrument response is 0.8 $imp\ h^{-1}$ per $Bq\ m^{-3}$. The maximum $^{222}Rn$ concentration that the instrument can measure is 150 $kBq\ m^{-3}$ within a time interval of one hour, whereas the rather low activity concentration of 30 $Bq\ m^{-3}$ is determined with a statistical accuracy equal to ±20% for a counting time of only one hour. The instrument was adjusted to record the data every 2 hours over the 24-hour measuring period. Therefore, we could detect any variations in pressure, temperature, and humidity (see Fig. 4). The data could be read in counts or $Bq\ m^{-3}$ and could be displayed on the instrument screen or could be transferred and displayed on a computer monitor. The instrument calculated the mean, maximum, and minimum radon concentration over the adjusted time intervals.

For the measurements of radon concentration in drinking water samples, a RADIM3W (Plch, 2008), commercial name, compact detector system was utilized. It monitors the temperature, relative humidity and pressure and corrects the results for the effect of humidity. Polyethylene bottles were used to collect the water. They were completely filled with water so that no headspace was presented. By the end of the sampling day, the measurements were conducted at the laboratory.

The instrument is shown schematically in Fig. 2. The collected water sample was placed in a cylindrical glass bottle with an optimal volume of 250 *mL*. The mouth of the bottle was covered by a filter, which protects the diffusion of radon from the tested water evaporation. An external pump was utilized to release the radon from water by bubbling (Plch, 2008). More than 99% of radon is released into the air by bubbling if the volume of the bubbled air is 10 times more than the volume of the water. The aquarium pump type Terratec Model 50, used in the Radim3W system, is modified so that the pump to suck the air from an added input and not from the open air. The pump is connected into a closed circuit in which the air circulates between the aerator and the closed detection chamber of Radim3W. For the bubble production a normal aquarium stone is used. The pump gives a flow rate of 0.8 $L\ min^{-1}$ in the open air. However, the resistance of the stone reduces the flow rate. It was found experimentally that there is no difference on the results if bubbling lasts 5 minutes (theoretical total volume of air equal to 4 *L)* or 10 minutes, but 3 minutes of bubbling is too short to release all radon from 200 *mL* of water. Therefore, after 5 minutes of bubbling most part of the radon is in the detection chamber (0.8 *L*) but some part is also in the tubing (volume 130 *mL*) and in the aerator.

The radon eventually is directed to the closed detection chamber. The radon detection principal is similar to the RADIM3A monitor. All the equipment was installed into a room with relatively low radon concentration in air (8.0±3.2 $Bq\ m^{-3}$). The calibration procedure of the RADIM3W detection system is similar to the RADIM3A detection system.

The instrument response is 20 counts per $Bq\ L^{-1}$ in a 30-min measurement. It can measure a maximum concentration of 3 $kBq\ L^{-1}$ in a 30-min time interval and a minimum concentration of 1.5 $Bq\ L^{-1}$ within a 30-min measuring time with a statistical accuracy of ±20%. The bubbling time period was 5-min and the measurement time period was 30 min.

### *2.2. Impulses and Radon Concentration*

For the correct interpretation of the results, it is useful to clarify the relationship between the measured number of impulses $N$ and the radon concentration $C_{Rn}$. This relationship will be derived for the hypothetical case where the activity of radon and its decay products in the detection system is equal to zero**.** Thus, at zero time, the high voltage is turned on and the instrument is brought into a space with radon concentration. Radon diffuses into the detection chamber and decays to yield its decay products. The monitor detects only α-particles and selects those impulses formed by the decay of $^{218}Po$, with half-time ($T_{1/2}$) equal to 3 minutes. The $^{218}Po$ is practically in equilibrium with radon after about 6 half-times (i.e., after about 20 minutes). Therefore, the increase in the overall activity deposited on the surface of the detector is dependent on the rate of radon diffusion into the chamber and the rate of $^{218}Po$ formation through the radon decay. Neglecting the diffusion, the increase in the measuring number of impulses N can be described by the following equation:

$$N = n_s \left( 1 - e^{-\lambda t} \right) t \tag{1}$$

where $\lambda = ln2/T_{1/2} \approx 0.23\ min^{-1}$ is the decay constant, $n_s$ is equal to the saturated counting rate of impulses *(in count/s)*, and $t$ is the counting time. Twenty minutes after the sudden increase in activity, the exponential term in Eq. (1) is equal to 0.01 and it yields $N \approx n_s\, t$.

In real measuring conditions, however, the monitor is placed in a space with an initial non-zero activity at zero time. Thus, the number of impulses N, measured over time T, can be calculates as:

$$N = n_s \int_0^T \left( 1 - e^{-\lambda t} \right) dt, \tag{2}$$

which, after integration, results in:

$$N = n_s \left[ T + \frac{1}{\lambda} \left( e^{-\lambda T} - 1 \right) \right] \tag{3}$$

After attaining equilibrium ($T >> \lambda^{-1} = 4.3\ min$), the exponential term ($e^{-\lambda T}$) in Eq. (3) becomes negligible, so that the total number of impulses is then equal to:

$$N = n_s \left( T - \frac{1}{\lambda} \right). \tag{4}$$

Thus, this initial "growth" period (equal to $\lambda^{-1} \simeq 4.3\ min$) has to be subtracted from the total measuring time to correctly determine the saturation activity ($n_s$) from the measured number of impulses $N$ via Eq. (4).

Note, that the above description is only approximate. Nonetheless, it is completely satisfactory as long as the measuring time $t$ is sufficiently long; i.e. $t >> \lambda^{-1} \simeq 4.3\, min$.

Hence, the radon concentration, $C_{Rn}$, can be calculated from the equilibrium saturated counting rate of impulses ($n_s$) according to the following relation:

$$C_{Rn} = \frac{n_s}{V\, \varepsilon_d\, d_s\, \varepsilon_s} \tag{5}$$

where the radon concentration is in $Bq\ m^{-3}$ or in $Bq\ L^{-1}$, $V$ is the volume of the chamber (in $m^3$ or $L$), $\varepsilon_d$ is the detection efficiency, $d_s$ is the probability of formation of positively charged radon decay products, and $\varepsilon_s$ is the collection efficiency of the detector.

From the detection principle, the detection efficiency is practically equal to 50%. Further, it is stated in the literature that about 80% and 87% of radon decay products are positively charged in the RADIM3A and RADIM3W detection systems, respectively. The RADIM3A and RADIM3W detectors volume $V$ are equal to 0.7 $L$ and 0.83 $L$, respectively. The collection efficiency of positively charged radon decay products is the unknown quantity, which depends sensitively on humidity. Therefore, the factor $k = \varepsilon_d\, d_s\, \varepsilon_s$ in the denominator of Eq. (5) is a critical quantity and represents the response function of the detector. It has been carefully studied as a function of the absolute humidity and adequately parameterized. The results are stored in the individual internal microprocessor of each instrument by the manufacturer (Plch, 2001; Plch, 2008). Thus, in real measuring conditions, the absolute humidity is calculated by the simultaneous measurements of the relative humidity and temperature provided by the sensors of the instruments within the relevant regions of interest (temperature between 0 $^0C$ and 40 $^0C$, and relative humidity between 5% and 95%). Hence, for each measuring time interval, the internal processor of the instrument automatically calculates, first, the actual value of the response function of the detector, and eventually, the radon concentration. The latter, in this way, is corrected for the effect of humidity. For instance, a change in the relative humidity from 50% to 90% decreases the RADIM3A sensitivity by about 6.5% and the RADIM3W sensitivity by about 27%.

From the study of the detectors response, it is estimated that the collection efficiency of the RADIM3A system is about 0.8 and for the RADIM3W system is about 0.4 (Plch, 2001; Plch, 2008).

*2.3. Site selection and counting*

A total of 108 airborne indoor radon concentration measurements were conducted to construct the radon map of the central part of the Nicosia district. Initially, the central part of the district was divided into 54 grids of 1 $km^2$ area each (Fig. 3a).

Two 24hr measurements (12 records per measurement) were obtained in each grid and the average value was recorded as the radon concentration value for the grid. The pressure, relative humidity, temperature and radon concentration were observed every 2 hours (sampling time) for each 24hr measurement, to identify possible diurnal variations (see Fig.

4). The measurements lasted 9 months (9/2008 – 5/2009) and the two measurements in each grid were conducted in different seasons of the year to observe possible seasonal variations. It should be noted here that time variations in the radon release is expected to occur when radon escapes from the soil and reaches the surface of a place, mainly depending on the geodynamics of the specific place. Thus, seasonal variations of the measured radon activity may occur due to the specific climatic conditions that govern in a certain location (see. e.g. Neville and Hultquist, 2008 and further references cited therein). In Cyprus, however, such effects are very small due to the fact that the radon activity is very low and the weather conditions are stable and good nearly throughout the whole year.

A Global Position System (GARMIN, 2007) was utilized to identify and report the coordinates of the measuring sites. The geographical coordinates, in decimal degrees, of the measuring sites are presented in Table 1. The projected geographical coordinates of the measuring sites were used to position the sites on a map. Figure 3b shows the map of the measuring sites. The labels of the measuring sites are in agreement with the grid numbers in Fig. 3a and Table 1. Two measurements were conducted in each grid (see Table 1). The map of Fig. 3b is based upon the map prepared by the Department of Lands and Surveys with the sanction of the Government of Cyprus.

Priority for the site selection was given to schools (40 measurements), public workplaces (15 workplaces), and old dwellings (53 measurements). Drought-free areas in the sites were selected to place the radon monitor, such as basements, away from doors and windows, to record the maximum radon concentration. The detectors were always placed at a height of approximately 1 meter.

The averaged outdoor radon activity concentration, measured in 4 sites of the district, was 4.1±0.7$Bq\ m^{-3}$. This value agreed with the corresponding outdoor radon activity concentration of 3.9±0.8 $Bq\ m^{-3}$, measured 7 years ago using the same method (Anastasiou et al., 2003). The outdoor radon concentration value was not subtracted from the results obtained in the indoor measurements.

A total of 24 measurements were conducted to determine the radon levels in the drinking water within the central part of the Nicosia district. The drinking water was collected from 16 schools and 8 houses. The addresses of these sites are presented in Table 2.

*2.4. Annual effective dose rates*

The annual effective dose $H_E$ in $Sv\,y^{-1}$, to a member of the population due to the inhalation of radon in indoor air, was calculated according to the following equation (UNSCEAR, 2000)

$$H_E = C_{Rn} \times F \times T \times K, \qquad (6)$$

where $C_{Rn}$ in $Bq\,m^{-3}$ is the measured mean radon activity concentration in air, $F=0.4$ is the indoor equilibrium factor between radon and its progeny, $T$ is the indoor exposure time in hours per year (assumed to be equal to $T= 0.8{\times}24\ h\ {\times}365.25\ days \cong 7013\ h\,y^{-1}$ (1) ), and $K=9\ nSv\,h^{-1}$ per $Bq\,m^{-3}$ is the conversion factor from the absorbed dose in air to the corresponding effective dose (UNSCEAR, 2000).

The annual effective dose due to the inhalation of radon, resulting from radon and its decay products in drinking water, was calculated as follows:

$$H_E = C_{Rn} \times 10^{-4} \times F \times T \times K, \qquad (7)$$

where $C_{Rn}$ in $Bq\,m^{-3}$ is the measured mean radon activity concentration in drinking water and the factor $10^{-4}$ is the air to water concentration ratio (UNSCEAR, 2000).

## 3. Results and discussion

Table 1 shows the indoor airborne radon concentration for the measurements (two measurements in each of the 54 grids), the average radon concentration in each grid and the associated annual effective doses rates. The mean radon concentration measurements range from 4.5–151.4 $Bq\,m^{-3}$ with a mean value of 20.6±13.2 $Bq\,m^{-3}$, which is in fair agreement with the average value of 19.3±14.7 $Bq\,m^{-3}$, derived from our previous survey measurements (Anastasiou et al., 2003). The latter were carried out over a 9 months time interval (September 2001 to May 2002), utilizing the same detectors in different seasonal periods, with different sampling times. Also, the results are in the general trend consistent with the first radon concentration measurements by means of passive detectors, CR-39 element, carried out over 6 months, December 1990 to May 1991, (Christofides and Christodoulides, 1993).

Figure 4 shows the pressure, relative humidity, temperature, and radon concentration from a typical 24hr measurement with sampling time of 2 hours using the RADIM3A detection system. The measurement corresponds to the geographical coordinates 33.3313 and 35.1272 at grid number 31, started on 16/11/2008 at 11:30 am. The error bars are solely of statistical origin. The pressure and relative humidity during the measurement is constant increasing the reliability of the measurement. The observed diurnal variation in the radon concentration (i.e., higher mean radon concentration values during the morning hours and lower mean radon concentration values during the evening hours) is due to a corresponding change in the air ventilation rate. This was also observed in our previous investigations (Anastasiou et al., 2003).

Figure 5 presents the indoor airborne radon concentrations for the 54 grids. The solid line represents the average radon concentration of the measurements (20.6 $Bq\ m^{-3}$) and the dashed line represents the corresponding worldwide average radon concentration value, 39 $Bq\ m^{-3}$, (UNSCEAR, 2000). Note that the average world value is a conventional indicator to compare the present data since there is great variability of radon data in natural environments. The average radon concentrations for the grids with numbers 4, 22, and 24 are above the world average value. In particular, one of the two measurements in each of the abovementioned grids has a radon concentration value higher than the worldwide average value. These three relatively high radon concentration measurements with numbers 4a, 22b, and 24a (see Table 1) were obtained from schools. Further measurements at the specific sites were necessary to verify and/or justify these values. Hence, the detectors were placed in the same and different rooms of the buildings. For example, regarding the measurement with number 4a, eight months later, the radon concentration in a different room of the building was 14.4±4.1 $Bq\ m^{-3}$ and in the same room, was 96.2±14.0 $Bq\ m^{-3}$. Only the same room presented these relatively high values due to a very low air change compared to the other rooms of the building. Moreover, the initial radon concentration value of 151.4 $Bq\ m^{-3}$ was recorded at the beginning of the school year, in September, where the room was closed during summer vacations, while the repeated measurement of 96.2 $Bq\ m^{-3}$ was recorded in April, by the end of the school year. The two-radon concentration values from the same room (151.4 $Bq\ m^{-3}$ and 96.2 $Bq\ m^{-3}$) and the radon concentration value from the other room (14.4 $Bq\ m^{-3}$ ) were taken into consideration in Fig. 8.

The annual effective dose rates for the measurements in the 54 grids are plotted in Fig. 6a. The solid line represents the average annual effective dose rate value of the measurements (0.52 $mSv\ y^{-1}$) and the dashed line represents the corresponding worldwide average value, 1 $mSv\ y^{-1}$ (UNSCEAR, 2000). The rates for the grids with numbers 4, 22, and 24 are higher than 1 $mSv\ y^{-1}$. This was discussed in the previous paragraph. Figure 6b shows the percentage frequency distribution of the annual effective dose rates of Fig. 3a. Only in 5.5% of the grids (3 out of 54), the annual effective dose rates are higher than our reference value of 1 $mSv\ y^{-1}$.

Table 2 presents the 24 selected sites for the drinking water measurements and the associated radon concentrations. The values range from 27.0-1083.9 $mBq\ L^{-1}$ with an overall mean value of 243.8±224.8 $mBq\ L^{-1}$). The results are plotted in Fig. 7. The solid line represents the average value of the measurements (243.8 $mBq\ L^{-1}$) and the dashed line represents the accepted radon level of 11000 $mBq\ L^{-1}$ proposed by USEPA (1999). The average value of these measurements is far below the above value. The calculated annual effective dose due to inhalation of radon from drinking water ranges from 0.07-2.74 $\mu Sv\ y^{-1}$ with a mean value of 0.0616±0.0567 $\mu Sv\ y^{-1}$.

The first radon map of the central part of the Nicosia district is constructed and presented in Fig. 8 using the official digital map prepared by the Department of Lands and Surveys with the sanction of the Government of Cyprus. The ArcMap 9.3.1 (ESRI, 2010) software package was utilized to insert the geographical coordinates of the measuring sites and the associated indoor airborne radon concentration values of Table 1. The geographical coordinates of the measuring sites for each grid were averaged and then projected on the map with dots. The dots are labeled with the grid number (see Fig. 3a). The dot-sizes are proportional to the grid radon concentration values according to the legend, whereas the actual values are given in Table 1.

## 4. Conclusions

A total of 108 indoor airborne measurements were conducted in 54 grids, of 1 $km^2$ area each, of the central part of the Nicosia district. The overall mean value of the radon concentrations and the associated mean value of the annual effective dose rates are by almost a factor of two lower than the worldwide average values. In three schools, with relatively high radon concentration values, the measurements were repeated in various rooms of the buildings, and it was decided that the initial radon concentration value was due to the very low air change rate compared to the other rooms of the building. It is worth emphasizing in this context that special care should be taken by the school's authorities to ensure that good air ventilation is achieved in the class and office rooms, before the opening of the schools after the long summer holidays period.

An important feature of all the cases studied is the fact that the indoor airborne radon concentration values were found to be below the corresponding upper limits defined by other European countries (EC, 1997; Synnott and Fenton, 2005).

Further, the mean value of the radon levels in the drinking water from 24 measurements is below the upper limit proposed by USEPA (1999). In addition, the digital radon map of the central part of the Nicosia district was constructed by means of the ArcMap 9.3.1 software package.

The present systematic investigations on radon activity concentration in indoor air and in water clearly demonstrate that there is no radon risk in the high-population density area of the Nicosia district. Both geological background and climatic conditions result in a decreased airborne radon levels in the central part of the Nicosia district and, more generally, in whole Cyprus.

## Acknowledgements

The authors would like to thank P. Demetriades and M. Tzortzis from the Radiation Protection and Control Services, of the Department of Labor Inspection, of the Cyprus Ministry of Labor and Social Insurances, for the fruitful cooperative work. We also thank the directors of the buildings and the residents of the dwellings, for their cooperation.

## References

Anastasiou, T., Tsertos, H., Christofides, S., Christodoulides G., 2003. Indoor radon concentration measurements in Cyprus using high-sensitivity portable detectors. J. Environ.. Radioact. 68, 159-169.

Christofides, S., Christodoulides, G., 1993. Airborne radon concentrations in Cypriot houses. Health Phys. 64 (4), 392-396.

EC, 1997. Radiation Protection 88.

ESRI, 2010. Geographical Information System ArcGIS 9.3.1 Instruction Manual, www.esri.com

GARMIN Ltd, 2007. GPSMAP 60CSx with sensors and maps Owner's manual.

Giammanco, S., Immè, G., Mangano, G., Morelli, D., Neri, M., 2009. Comparison between different methodologies for detecting Radon in soil along an active fault: the case of the Pernicana fault system, Mt. Etna. Appl. Radiat. and Isot., 67(1), 178-185, doi: 10.1016/j.apradiso.2008.09.007.

HHS, 2010. 11$^{th}$ Report on Carcinogens, National Toxicology Program, U.S. Department of Health and Human Services:

http://ntp.niehs.nih.gov/INDEXA5E1.HTM?objectid=32BA9724-F1F6-975E7FCE50709CB4C932.

Hopke, P.K., 1989. Use of electrostatic collection of $^{218}$Po for measuring radon. Health Physics 57, 39 -42.

IAEA, 2003. Radiation Protection against Radon in Workplaces other than Mines, www-pub.iaea.org/MTCD/publications/PDF/Pub1168_web.pdf

ICRP, 2009. ICRP Statement on Radon, www.icrp.org.

Michael, F., Parpottas, Y., Tsertos, H., 2011. Gamma radiation measurements and dose rates in commonly used building materials in Cyprus. Radiat. Prot. Dosim., 142, Issue: 2-4, 282-291, doi: 10.1093/rpd/ncq193.

Neri, M., Giammanco, S., Ferrera, E., Patanè, G., Zanon, V., 2011. Spatial distribution of soil radon as a tool to recognize active faulting on an active volcano: the example of Mt. Etna (Italy). J. Environ. Radioact., 102, 863-870, doi: 10.1016/j.jenvrad.2011.05.002.

Neville, J., D., Hultquist, D., J., 2008. Seasonal radon variations in Utah testing results: short-term test results within 10% of the EPA threshold (4.0 pci/l) should be repeated in a different season. Proceedings of the American Association of Radon Scientists and Technologists 2008 International Symposium Las Vegas NV, September 14-17, 2008.

Plch, J., 2001. Radon monitor Radim3, Instruction Manual, S.K. Neumanna, 18200 Prague 8.

Plch, J., 2008. Radon monitor Radim3W, Instruction Manual, 18200 Prague 8.

Svoukis, E., Tsertos, H., 2007. Indoor and outdoor in situ high-resolution gamma radiation measurements in urban areas of Cyprus. Radiat. Protect. Dosim. 123 (3), 384-390.

Synnott, H., Fenton, D., 2005. An Evaluation of Radon Mapping Techniques in Europe, Radiological Protection Institute of Ireland.

Tzortzis, M., Tsertos, H., Christofides, S., Christodoulides, G., 2003. Gamma-ray measurements of naturally occurring radioactive samples from Cyprus characteristic geological rocks. Radiat. Meas. 37, 221-229.

Tzortzis, M., Tsertos, H., Christofides, S., Christodoulides, G., 2003. Gamma radiation measurements and dose rates in commercially-used natural tiling rocks (granites). J. Environ. Radioact. 70, 223-235.

Tzortzis, M., Svoukis, E., Tsertos, H., 2004. A comprehensive study of natural gamma radioactivity levels and associated dose rates from surface soils in Cyprus. Radiat. Prot. Dosim. 109 (3), 217-224.

Tzortzis, M., Tsertos, H., 2004. Determination of thorium, uranium and potassium elemental concentrations in surface soils in Cyprus, J. Environ. Radioact. 77 (3), 325-338.

UNSCEAR, 2000. Sources and Effects of Ionising Radiation, Report to General Assembly with Scientific Annexes, United Nations, New York.

USEPA, 1999. Office of ground water and drinking water rule: Technical Fact Sheet EPA 815-F-99-006, Washington DC, http://www.epa.gov/safewater/radon/fact.html.

WHO 2005. Fact Sheet No. 291.

## TABLES

Table 1

The geographical coordinates of the two measurements in each of the 54 grids, the associated minimum, maximum and mean indoor airborne radon concentration for each measurement, the average radon concentration for each grid, and the corresponding annual effective dose rates.

| Grid Number | Coordinates | | Radon Concentration ($Bq\ m^{-3}$) | | | | | | Annual Effective Dose Rates ($mSv\ y^{-1}$) | |
|---|---|---|---|---|---|---|---|---|---|---|
| | Longitude | Latitude | Min | Max | Avg. | Stat. Error | Grid Avg. | Stat. Error | Rates | Stat. Error |
| 1 | 33.3859 | 35.1872 | 5.4 | 14.3 | 10.5 | 0.9 | 20.0 | 1.0 | 0.503 | 0.026 |
| | 33.3863 | 35.1865 | 23.7 | 40.2 | 29.4 | 1.9 | | | | |
| 2 | 33.3889 | 35.1841 | 3.4 | 22.5 | 13.3 | 6.1 | 13.0 | 4.1 | 0.328 | 0.103 |
| | 33.3881 | 35.1829 | 4.2 | 24.9 | 12.7 | 5.5 | | | | |
| 3 | 33.3203 | 35.1710 | 4.2 | 13.1 | 7.2 | 1.0 | 11.8 | 1.0 | 0.297 | 0.026 |
| | 33.3217 | 35.1723 | 11.9 | 27.5 | 16.3 | 1.8 | | | | |
| 4 | 33.3313 | 35.1750 | 105.0 | 196.9 | 151.4 | 8.2 | 86.4 | 4.2 | 2.180 | 0.106 |
| | 33.3597 | 35.1682 | 15.2 | 29.4 | 21.4 | 1.9 | | | | |
| 5 | 33.4124 | 35.1523 | 4.2 | 12.4 | 8.3 | 0.8 | 12.3 | 1.1 | 0.310 | 0.029 |
| | 33.3486 | 35.1727 | 5.9 | 29.0 | 16.3 | 2.2 | | | | |
| 6 | 33.3616 | 35.1694 | 8.5 | 16.8 | 10.3 | 1.4 | 11.5 | 1.2 | 0.290 | 0.030 |
| | 33.3486 | 35.1727 | 4.2 | 21.5 | 12.7 | 1.9 | | | | |
| 7 | 33.3735 | 35.1772 | 5.3 | 22.3 | 12.5 | 1.7 | 23.1 | 2.6 | 0.583 | 0.065 |
| | 33.3707 | 35.1757 | 12.8 | 55.7 | 33.7 | 4.9 | | | | |
| 8 | 33.3796 | 35.1707 | 9.6 | 26.2 | 17.1 | 2.1 | 14.2 | 1.3 | 0.357 | 0.033 |
| | 33.3785 | 35.1849 | 7.8 | 20.1 | 11.2 | 1.6 | | | | |
| 9 | 33.3957 | 35.1728 | 8.0 | 26.3 | 18.6 | 6.1 | 16.6 | 3.9 | 0.419 | 0.098 |
| | 33.3958 | 35.1722 | 4.9 | 21.1 | 14.6 | 4.8 | | | | |
| 10 | 33.3050 | 35.1551 | 4.7 | 14.1 | 7.8 | 0.8 | 11.6 | 1.1 | 0.291 | 0.027 |
| | 33.3080 | 35.1787 | 11.2 | 32.1 | 15.3 | 2.0 | | | | |
| 11 | 33.3126 | 35.1614 | 6.9 | 28.9 | 16.1 | 2.5 | 20.9 | 3.1 | 0.526 | 0.078 |
| | 33.3129 | 35.1557 | 5.7 | 55.5 | 25.6 | 5.7 | | | | |
| 12 | 33.3280 | 35.1678 | 4.5 | 13.0 | 8.6 | 0.8 | 18.9 | 1.1 | 0.476 | 0.027 |
| | 33.3323 | 35.1616 | 22.3 | 40.7 | 29.1 | 2.0 | | | | |
| 13 | 33.3451 | 35.1625 | 21.3 | 44.9 | 28.5 | 2.5 | 27.8 | 1.5 | 0.702 | 0.039 |
| | 33.3487 | 35.1647 | 20.2 | 38.1 | 27.1 | 1.9 | | | | |
| 14 | 33.3574 | 35.1639 | 12.1 | 47.1 | 30.3 | 11.2 | 30.0 | 7.2 | 0.756 | 0.183 |
| | 33.3523 | 35.1611 | 17.3 | 43.7 | 29.6 | 9.2 | | | | |
| 15 | 33.3738 | 35.1585 | 1.8 | 8.9 | 5.5 | 0.7 | 7.2 | 0.5 | 0.180 | 0.014 |
| | 33.3733 | 35.1619 | 5.2 | 14.1 | 8.8 | 0.9 | | | | |
| 16 | 33.3871 | 35.1563 | 9.5 | 23.4 | 14.9 | 1.5 | 17.5 | 2.15 | 0.440 | 0.054 |
| | 33.3774 | 35.1601 | 6.5 | 58.6 | 20.0 | 4.0 | | | | |
| 17 | 33.3919 | 35.1564 | 1.8 | 16.9 | 11.1 | 1.2 | 9.4 | 0.9 | 0.236 | 0.022 |
| | 33.3901 | 35.1578 | 1.7 | 14.4 | 7.6 | 1.3 | | | | |
| 18 | 33.3987 | 35.1612 | 10.7 | 31.0 | 16.7 | 5.8 | 14.3 | 3.6 | 0.361 | 0.092 |
| | 33.3898 | 35.1602 | 6.2 | 21.3 | 11.9 | 4.4 | | | | |
| 19 | 33.3040 | 35.1512 | 6.4 | 25.4 | 13.8 | 2.0 | 14.6 | 1.3 | 0.368 | 0.033 |
| | 33.3198 | 35.1611 | 8.2 | 22.0 | 15.4 | 1.7 | | | | |
| 20 | 33.3160 | 35.1411 | 14.0 | 42.8 | 26.9 | 2.4 | 25.6 | 1.6 | 0.646 | 0.04 |
| | 33.3180 | 35.1423 | 16.3 | 35.8 | 24.3 | 2.1 | | | | |
| 21 | 33.3239 | 35.1437 | 5.8 | 21.4 | 12.8 | 1.6 | 12.2 | 1.1 | 0.308 | 0.028 |
| | 33.3134 | 35.1441 | 7.9 | 18.7 | 11.6 | 1.6 | | | | |
| 22 | 33.3433 | 35.1452 | 4.5 | 56.4 | 30.9 | 5.6 | 49.3 | 4.25 | 1.244 | 0.1075 |
| | 33.3405 | 35.1377 | 35.8 | 93.9 | 67.7 | 6.4 | | | | |
| 23 | 33.3604 | 35.1522 | 4.2 | 9.0 | 5.1 | 0.7 | 7.2 | 0.5 | 0.180 | 0.0145 |
| | 33.3580 | 35.1450 | 6.3 | 16.3 | 9.2 | 0.9 | | | | |
| 24 | 33.3709 | 35.1404 | 57.3 | 136.2 | 89.0 | 12.6 | 59.5 | 6.6 | 1.500 | 0.168 |
| | 33.3642 | 35.1464 | 9.3 | 56.9 | 29.9 | 4.3 | | | | |
| 25 | 33.3785 | 35.1465 | 6.4 | 27.1 | 14.0 | 1.9 | 12.7 | 1.2 | 0.320 | 0.030 |
| | 33.3791 | 35.1521 | 4.8 | 19.5 | 11.4 | 1.5 | | | | |
| 26 | 33.3919 | 35.1486 | 3.1 | 16.4 | 9.2 | 1.0 | 10.9 | 0.9 | 0.274 | 0.022 |
| | 33.3992 | 35.1449 | 4.7 | 16.7 | 12.5 | 1.5 | | | | |
| 27 | 33.4121 | 35.1525 | 5.1 | 25.3 | 20.8 | 1.4 | 21.4 | 1.2 | 0.540 | 0.031 |
| | 33.4098 | 35.1499 | 6.2 | 28.5 | 22.0 | 2.0 | | | | |
| 28 | 33.3402 | 35.1562 | 2.3 | 17.1 | 11.2 | 1.6 | 17.9 | 1.3 | 0.452 | 0.033 |
| | 33.3406 | 35.1575 | 6.4 | 28.9 | 24.6 | 2.1 | | | | |
| 29 | 33.4122 | 35.1529 | 5.4 | 24.7 | 17.4 | 2.3 | 18.6 | 1.9 | 0.469 | 0.047 |
| | 33.3106 | 35.1211 | 6.6 | 26.2 | 19.8 | 3.0 | | | | |
| 30 | 33.3185 | 35.1366 | 4.1 | 15.5 | 8.3 | 1.0 | 6.4 | 0.5 | 0.162 | 0.013 |
| | 33.3193 | 35.1279 | 4.1 | 7.3 | 4.5 | 0.4 | | | | |
| 31 | 33.3313 | 35.1272 | 18.2 | 38.8 | 27.8 | 2.1 | 22.8 | 1.9 | 0.575 | 0.049 |
| | 33.3314 | 35.1277 | 10.1 | 35.1 | 17.8 | 3.3 | | | | |
| 32 | 33.3418 | 35.1348 | 8.6 | 21.5 | 14.0 | 1.4 | 16.0 | 1.4 | 0.403 | 0.036 |
| | 33.3444 | 35.1271 | 9.1 | 28.3 | 17.9 | 2.5 | | | | |
| 33 | 33.4006 | 35.1550 | 1.8 | 10.8 | 5.4 | 0.7 | 8.0 | 0.7 | 0.201 | 0.017 |
| | 33.3594 | 35.1302 | 4.3 | 19.2 | 10.5 | 1.2 | | | | |
| 34 | 33.3755 | 35.1274 | 1.1 | 11.7 | 7.6 | 1.1 | 10.2 | 0.7 | 0.256 | 0.018 |
| | 33.3721 | 35.1301 | 2.1 | 16.3 | 12.7 | 0.9 | | | | |
| 35 | 33.3367 | 35.1287 | 4.1 | 12.6 | 9.5 | 0.7 | 15.6 | 0.6 | 0.394 | 0.016 |
| | 33.3372 | 35.1296 | 6.9 | 29.1 | 21.7 | 1.1 | | | | |
| 36 | 33.3058 | 35.1105 | 1.3 | 13.4 | 7.4 | 0.6 | 7.3 | 0.6 | 0.183 | 0.014 |
| | 33.3048 | 35.1090 | 2.7 | 10.2 | 7.1 | 1.0 | | | | |
| 37 | 33.3069 | 35.1128 | 6.2 | 31.1 | 24.3 | 2.1 | 25.1 | 1.2 | 0.632 | 0.031 |
| | 33.3118 | 35.1158 | 5.7 | 34.6 | 25.8 | 1.3 | | | | |
| 38 | 33.3208 | 35.1130 | 3.8 | 31.9 | 26.5 | 2.6 | 25.7 | 1.3 | 0.649 | 0.034 |
| | 33.3309 | 35.1135 | 6.2 | 29.8 | 24.9 | 0.8 | | | | |
| 39 | 33.3387 | 35.1141 | 5.7 | 33.4 | 27.8 | 2.4 | 25.5 | 1.5 | 0.642 | 0.038 |
| | 33.3390 | 35.1140 | 4.3 | 28.1 | 23.1 | 1.8 | | | | |
| 40 | 33.3645 | 35.1057 | 5.6 | 28.8 | 24.6 | 2.1 | 23.9 | 1.4 | 0.603 | 0.035 |
| | 33.3581 | 35.1052 | 4.8 | 27.9 | 23.2 | 1.8 | | | | |
| 41 | 33.3751 | 35.1066 | 16.9 | 38.9 | 28.0 | 7.4 | 26.2 | 5.1 | 0.660 | 0.128 |
| | 33.3760 | 35.1068 | 13.2 | 34.4 | 24.3 | 7.0 | | | | |
| 42 | 33.3836 | 35.1121 | 8.4 | 20.1 | 11.2 | 1.6 | 17.9 | 1.3 | 0.452 | 0.033 |
| | 33.3829 | 35.1097 | 18.2 | 40.9 | 24.6 | 2.1 | | | | |
| 43 | 33.3998 | 35.1073 | 3.0 | 33.0 | 14.3 | 9.7 | 20.7 | 5.7 | 0.522 | 0.143 |
| | 33.3997 | 35.1084 | 16.3 | 35.6 | 27.1 | 5.9 | | | | |
| 44 | 33.4084 | 35.1036 | 5.3 | 14.8 | 10.9 | 0.7 | 11.2 | 0.6 | 0.281 | 0.016 |
| | 33.4082 | 35.1022 | 6.9 | 18.6 | 11.4 | 1.1 | | | | |
| 45 | 33.4165 | 35.1047 | 12.3 | 93.9 | 46.1 | 7.9 | 33.9 | 4.0 | 0.854 | 0.101 |
| | 33.4201 | 35.1063 | 11.4 | 29.8 | 21.6 | 1.2 | | | | |
| 46 | 33.3577 | 35.1227 | 6.5 | 27.7 | 17.1 | 7.4 | 21.0 | 5.3 | 0.529 | 0.134 |
| | 33.3600 | 35.1206 | 12.0 | 37.7 | 24.8 | 7.6 | | | | |
| 47 | 33.3681 | 35.1025 | 15.7 | 35.1 | 21.3 | 4.2 | 17.2 | 5.5 | 0.433 | 0.140 |
| | 33.3705 | 35.0988 | 3.1 | 28.6 | 13.0 | 10.3 | | | | |
| 48 | 33.3854 | 35.0969 | 15.5 | 35.1 | 26.3 | 5.7 | 21.1 | 3.4 | 0.531 | 0.087 |
| | 33.3822 | 35.0967 | 9.9 | 23.5 | 15.8 | 3.9 | | | | |
| 49 | 33.3901 | 35.0964 | 17.3 | 31.6 | 22.1 | 4.3 | 21.4 | 2.3 | 0.539 | 0.059 |
| | 33.3906 | 35.0968 | 16.4 | 29.7 | 20.6 | 1.9 | | | | |
| 50 | 33.4062 | 35.1014 | 19.0 | 34.8 | 27.2 | 1.9 | 25.7 | 3.0 | 0.649 | 0.077 |
| | 33.3997 | 35.1002 | 13.4 | 33.1 | 24.2 | 5.8 | | | | |
| 51 | 33.4250 | 35.1069 | 6.4 | 55.8 | 24.5 | 4.5 | 24.2 | 3.3 | 0.609 | 0.083 |
| | 33.4242 | 35.1075 | 13.1 | 54.8 | 23.8 | 4.8 | | | | |
| 52 | 33.4301 | 35.1079 | 9.9 | 50.2 | 23.2 | 6.7 | 23.9 | 3.7 | 0.602 | 0.093 |
| | 33.4397 | 35.1083 | 18.2 | 46.7 | 24.5 | 3.1 | | | | |
| 53 | 33.4290 | 35.1064 | 9.3 | 24.9 | 17.8 | 2.0 | 20.7 | 2.6 | 0.522 | 0.065 |
| | 33.4278 | 35.0999 | 10.4 | 38.9 | 23.6 | 4.8 | | | | |
| 54 | 33.4307 | 35.1002 | 14.8 | 44.2 | 24.6 | 2.8 | 24.3 | 1.8 | 0.612 | 0.046 |
| | 33.4311 | 35.1096 | 18.7 | 33.2 | 23.9 | 2.4 | | | | |
| Arithmetic Mean ± Standard Deviation (A.M. ± S.D.): | | | | | | | 20.6 ± 13.2 | | 0.519 ± 0.333 | |
| Geometric Mean ± Standard Deviation (G.M. ± S.D.): | | | | | | | 17.9 ± 13.2 | | 0.451 ± 0.333 | |

Table 2
The selected sites for the drinking water measurements, the calculated radon concentrations, $C_{Rn}$, and the associated statistical errors.

| Site Number | Site Name | Radon Concentration ($mBq\ L^{-1}$) | |
|---|---|---|---|
| | | $C_{Rn}$ | $\Delta C_{Rn}$ |
| 1 | Dasoupoli Primary School | 76.5 | 18.5 |
| 2 | Aglantzia C’ Primary School | 27.0 | 11.0 |
| 3 | Aglantzia High School | 143.9 | 25.4 |
| 4 | House at Aglantzia (Europis 3) | 89.9 | 20.1 |
| 5 | House at Aglantzia (Amanthoutos 5) | 481.2 | 46.5 |
| 6 | Kaimakli A’ Primary School | 292.3 | 36.3 |
| 7 | House at Kaimakli (M. Alexandrou 17) | 215.9 | 31.2 |
| 8 | Lakatamia C’ Primary School | 40.5 | 13.5 |
| 9 | House at Lakatamia (Oresti 3) | 54.0 | 15.6 |
| 10 | Latsia A’ Primary School | 72.0 | 18.0 |
| 11 | Latsia High School | 242.9 | 33.0 |
| 12 | House at Tseri (Analionta 70) | 148.4 | 25.8 |
| 13 | Egkomi High School | 431.8 | 44.1 |
| 14 | Likavitos Primary School | 449.7 | 45.0 |
| 15 | Ag. Omologites Primary School | 206.9 | 30.5 |
| 16 | Agia Marina Primary School | 166.4 | 27.4 |
| 17 | Ap. Barnabas Primary School | 409.3 | 42.9 |
| 18 | Ag. Dometios A’ Primary School | 202.4 | 30.2 |
| 19 | Dali Primary School | 386.8 | 41.7 |
| 20 | Dali High School | 107.9 | 22.0 |
| 21 | House at Dali (Thessalonikis 4) | 229.4 | 32.1 |
| 22 | Pera Chorio Primary School | 134.9 | 24.6 |
| 23 | House at Pera Chorio (Gr. Afxentiou 31) | 1083.9 | 69.8 |
| 24 | House at Nisou (Bouboulinas 3) | 157.4 | 26.6 |
| A.M. ± S.D.: | | 243.8 ± 224.8 | |
| G.M. ± S.D.: | | 173.6 ± 224.8 | |

**FIGURES**

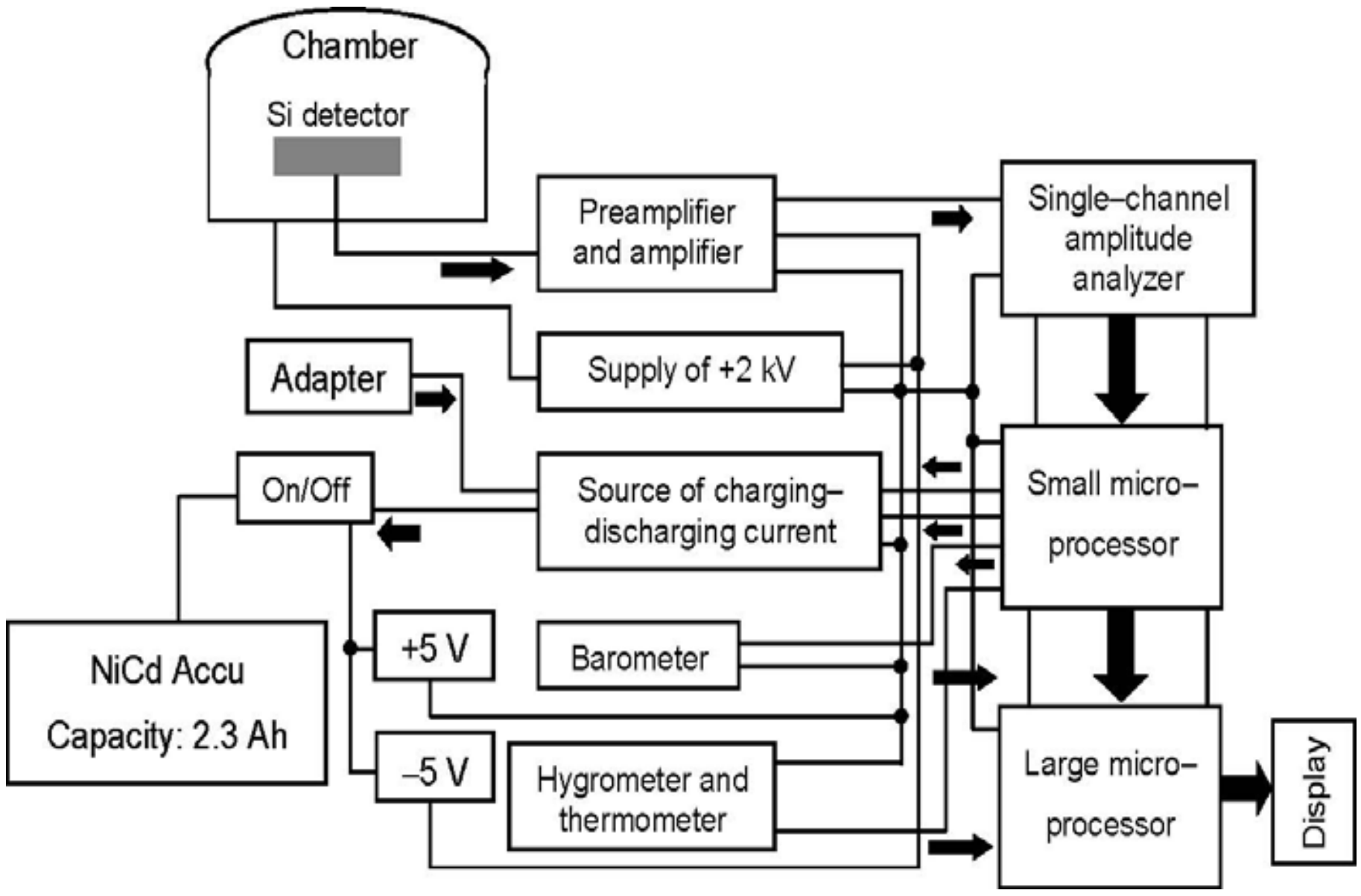

Fig. 1. Block diagram of the RADIM3A radon detection system.

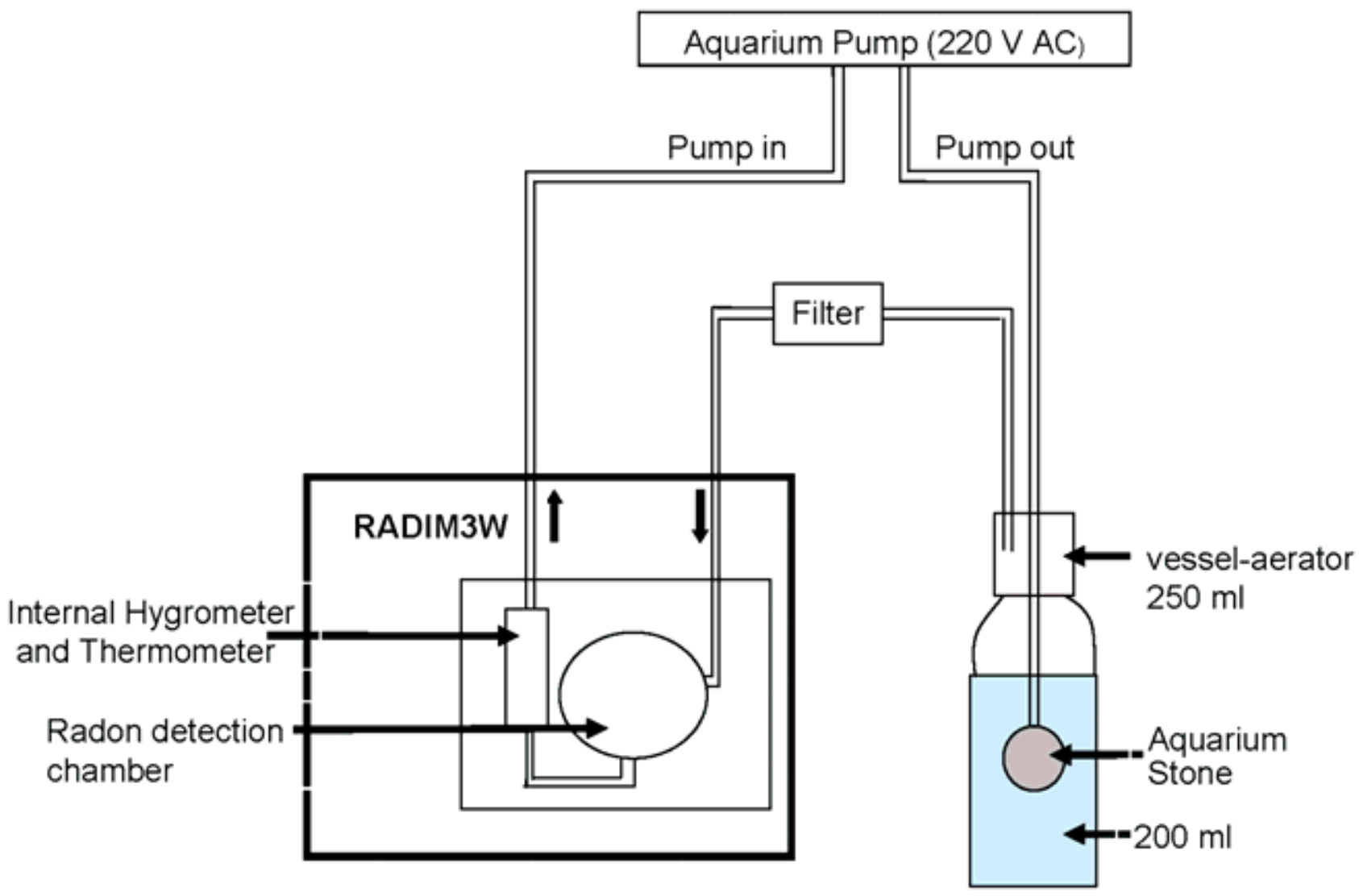

Fig. 2. Block diagram of the RADIM3W detection system.

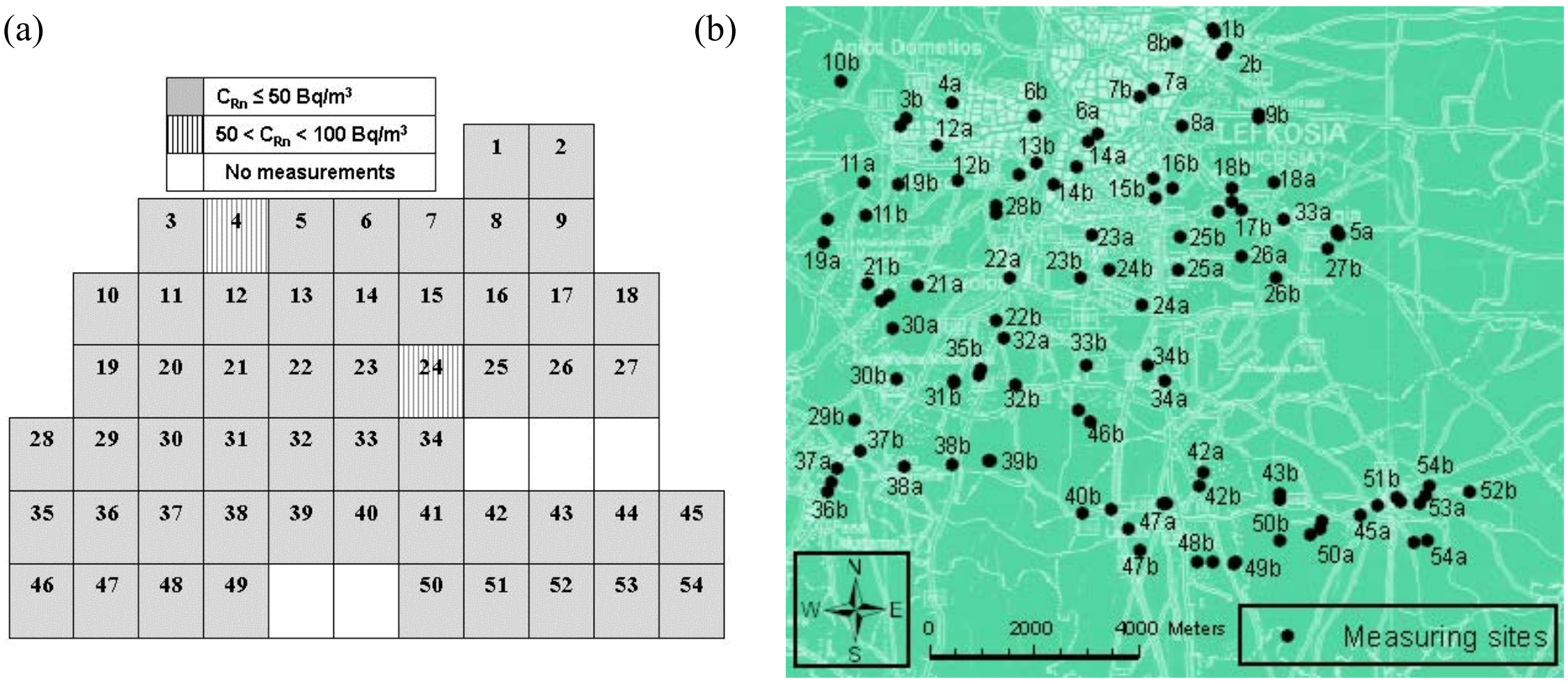

Fig. 3a. The 54 grids, with an area of 1 km[2] each, of the central part of the Nicosia district. The color of each grid represents the average radon concentration measured in each grid ($C_{Rn\ grid}$) according to the figure legend and Table 1. Fig. 3b. The digital map of the central part of the Nicosia district. It is based upon the map prepared by the Department of Lands and Surveys with the sanction of the Government of Cyprus. State Copyright Reserved. The measuring sites are indicated with dots and they are labeled according to the number of the measured grid in Fig. 1a. Two measurements were conducted in each grid (see Table 1).

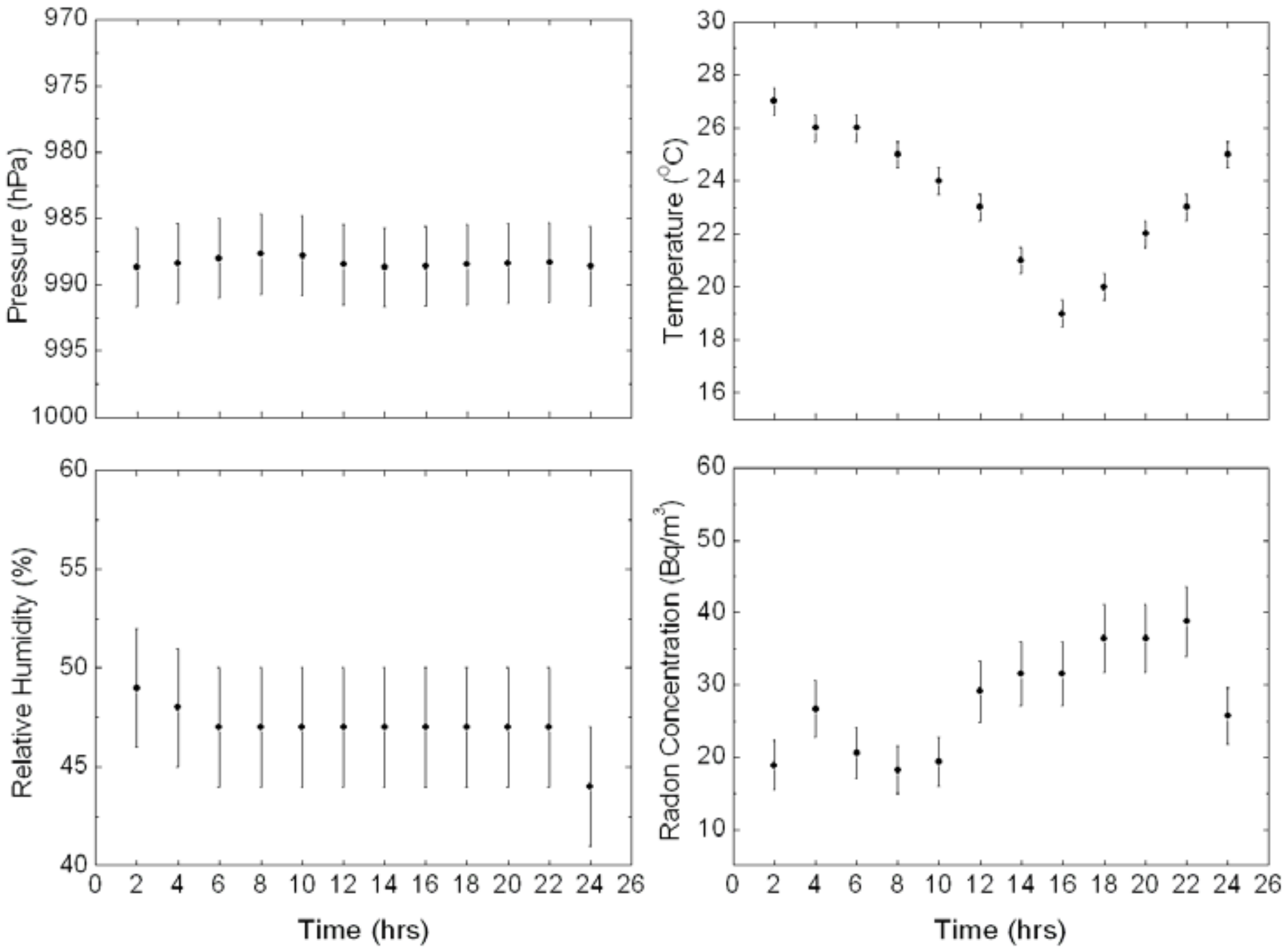

Fig. 4. The pressure, relative humidity, temperature and radon concentration from a typical 24hr measurement with sampling time of 2 hours using the RADIM3A detection system. The measurement corresponds to the geographical coordinates 33.3313 and 35.1272 at grid number 31, started on 16/11/2008 at 11:30 am. The error bars are solely of statistical origin.

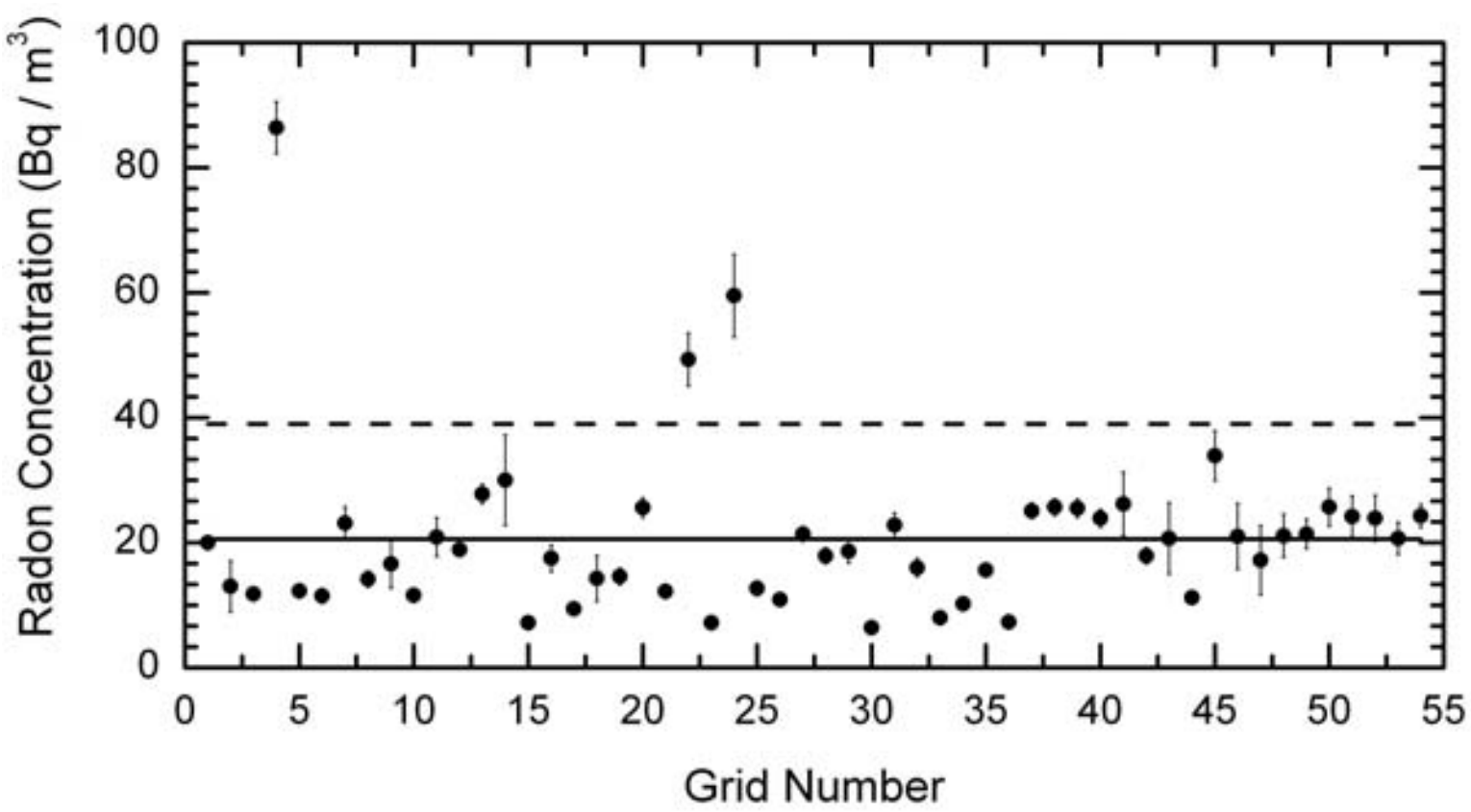

Fig. 5. The indoor airborne radon concentration values for the 54 grids of the central part of the Nicosia district. The solid line represents the average radon concentration of the measurements of 20.6 $Bq\ m^{-3}$ and the dashed line represents the corresponding worldwide average radon concentration value of 39 $Bq\ m^{-3}$ (UNSCEAR, 2000).

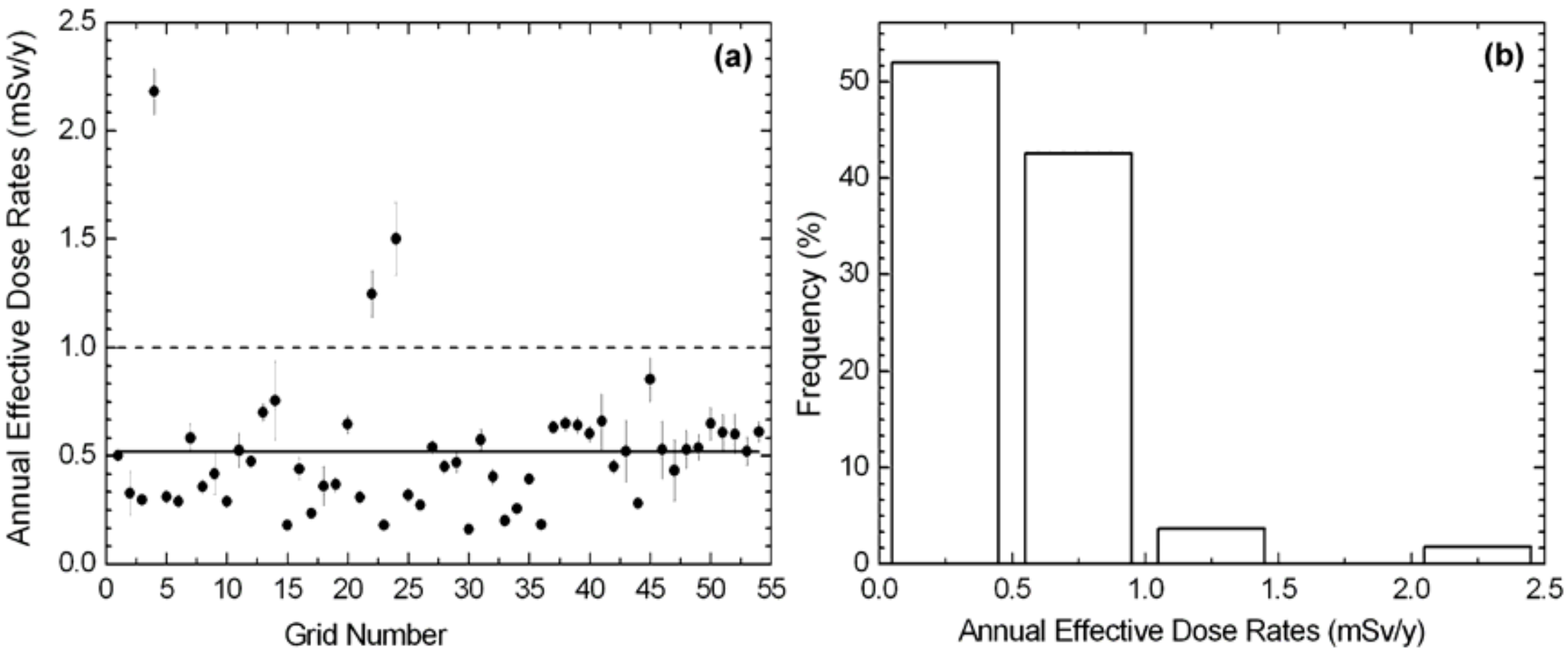

Fig. 6a. The annual effective dose rates of the measurements in the 54 grids. The solid line represents the average annual effective dose rate of 0.52 $mSv\ y^{-1}$ and the dashed line represents the corresponding worldwide average radon value of 1 $mSv\ y^{-}$ (UNSCEAR, 2000). Fig. 6b. The percentage frequency distribution of the annual effective dose rates of Fig. 3a.

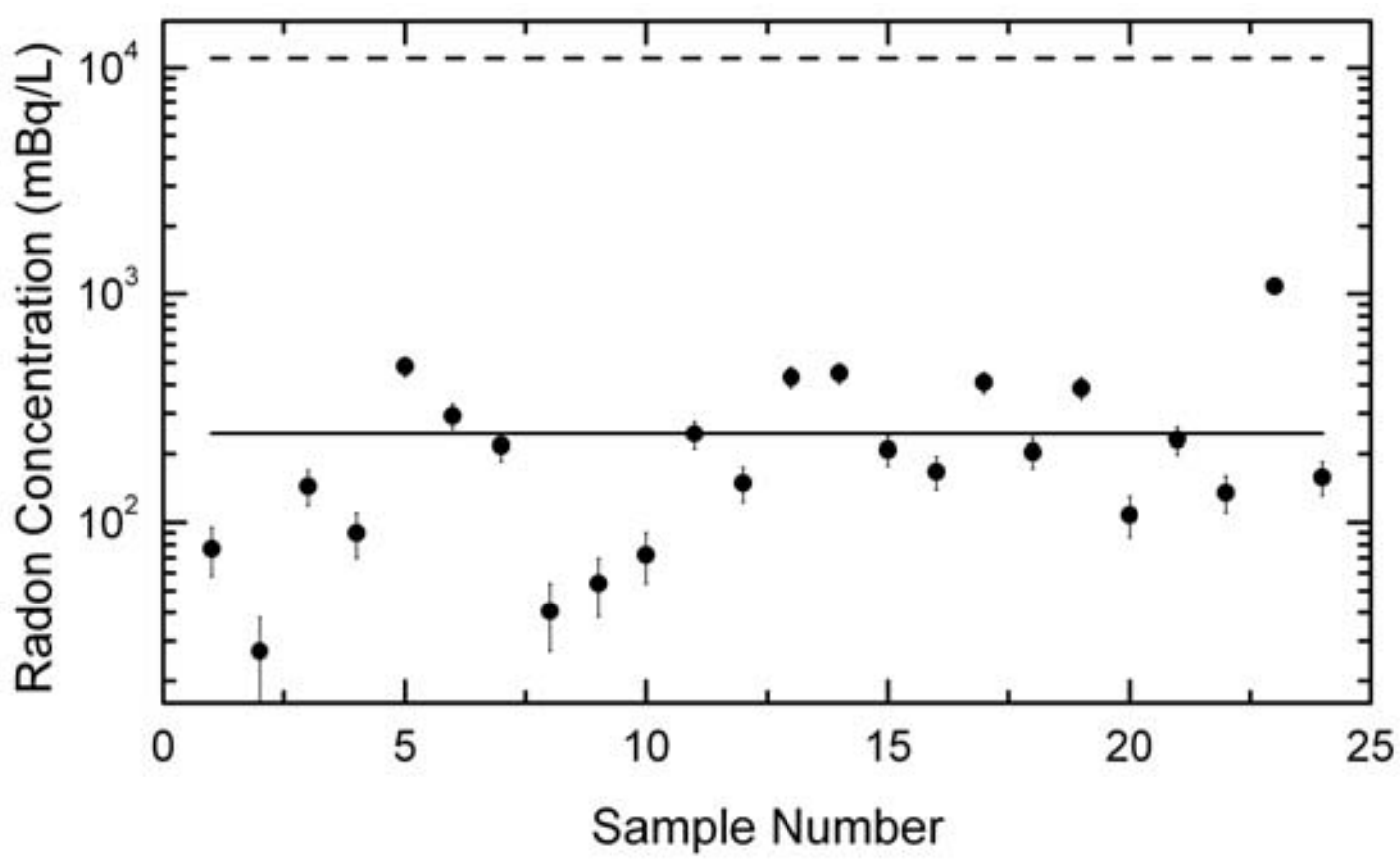

Fig. 7. The radon concentration levels in drinking water for these measurements. The solid line represents the average value of the measurements of 243.8 $mBq\ L^{-1}$ and the dashed line represents the upper limit of 11000 $mBq\ L^{-1}$ proposed by USEPA (1999).

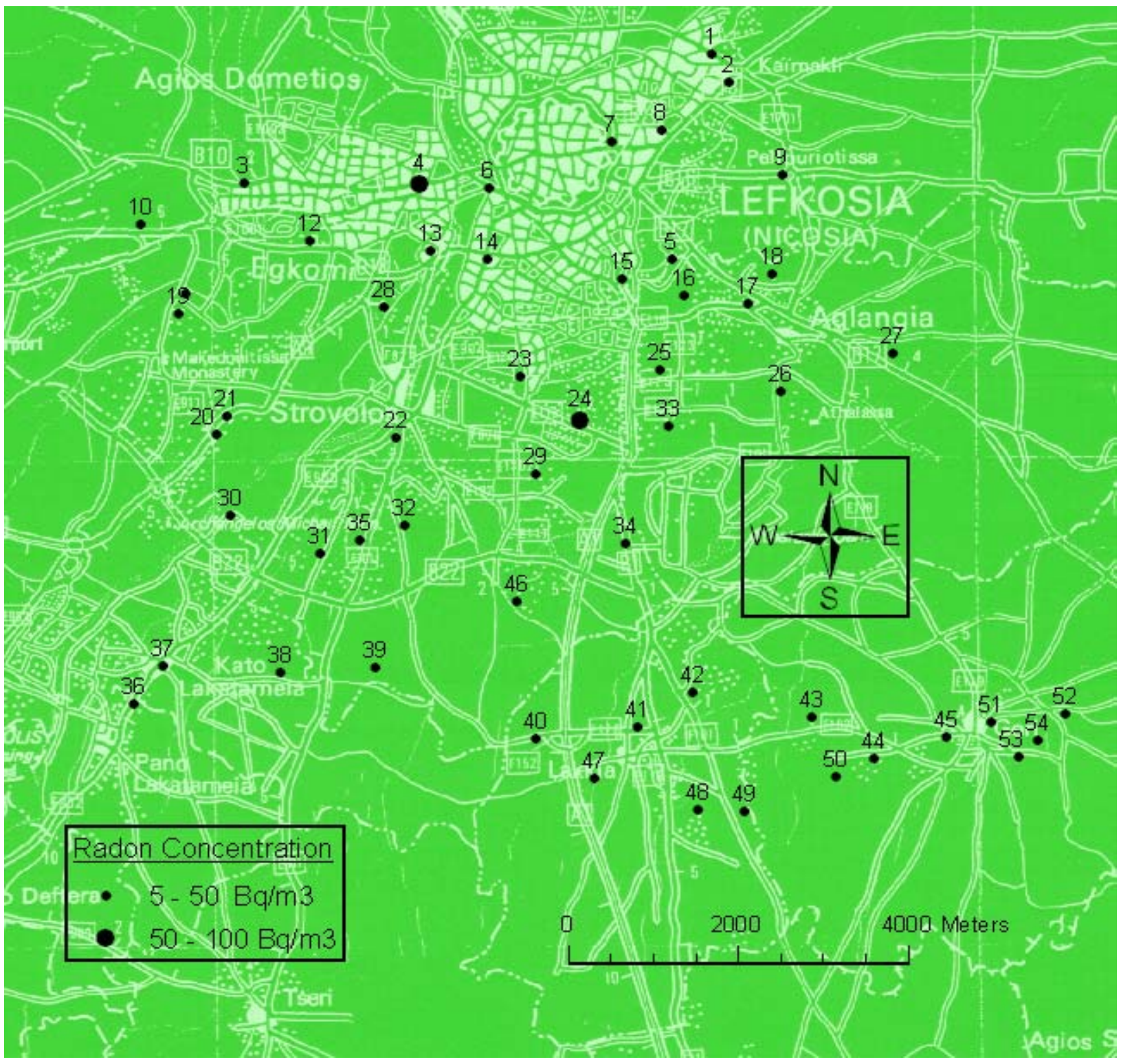

Fig. 8. The radon map of the central part of the Nicosia district. This map is based upon the official map prepared by the Department of Lands and Surveys with the sanction of the Government of Cyprus. State Copyright Reserved. The geographical coordinates of the measuring sites for each grid were averaged and then projected on the map with dots. The dots are labeled with the grid number. The dot-sizes are proportional to the grid radon concentration values according to the legend, whereas the actual values are given in Table 1.